# Individual GRB sensitivity of a cubic-kilometer deep-sea neutrino telescope KM3NeT

D. Dornic[1]* and G. Lelaizant

*Centre de Physique des Particules de Marseille, Faculte des Sciences de Luminy, 163 Avenue de Luminy, 13288 Marseille Cedex 09, France*

Representing the KM3NeT consortium

**Abstract**

Gamma-ray bursts (GRB) are powerful and highly variable sources of gamma rays that indicate the existence of cosmic particle accelerators. Under the assumption of hadronic acceleration in the jet, the expected neutrino energy spectrum is derived according to the intrinsic fireball model parameters and to the observed electromagnetic data of GRBs measured with ground-based and satellite observations. Using the performance characteristics of a cubic-kilometre scale neutrino detector placed in the Mediterranean Sea, the number of events is calculated individually for all the GRBs having a known redshift below the horizon of this detector. The good angular resolution of this detector and the narrow time windows around the GRB detection time allow suppression of almost all the atmospheric neutrino background. From the SWIFT GRB catalogue, we have derived the mean characteristics of a burst in order to be detected as an individual point source by a cubic-kilometre detector.

[1] Communicating author. Tel: +33 (0) 4 91 82 72 00; Fax: +33 (0) 4 91 82 72 89; E-mail: dornic@cppm.in2p3.fr

## 1. Introduction

The detection of gamma ray photons up to several GeV [1] shows that GRBs are very powerful particle accelerators. Assuming that protons and nuclei are accelerated along with electrons, GRBs are very promising sources of ultra high energy cosmic rays (UHECR) and high energy (HE) neutrinos. The small difference in arrival time between photons and neutrinos allows for very efficient detection by under-sea or under-ice telescopes. The KM3NeT consortium [2] is currently working on the technical design of a future Mediterranean neutrino telescope, which will have an instrumented volume of at least 1 km$^3$ and an angular resolution better than 0.2°. A possible design and its performance can be found in [3].

Several models predict the emission of high energy neutrinos close in time with photons. These correlations have not, so far, been detected. The detection of these neutrinos would provide some important constraints on the physics of GRBs and the CR acceleration mechanism in the jet.

In the standard Fireball model [4], the GRB gamma ray comes from the synchrotron radiation of electrons accelerated by diffuse shock acceleration in the internal shocks in the jet. HE neutrinos are produced by photonuclear interaction of the protons accelerated in the internal shocks with the observed gamma-rays. The accelerated proton flux follows a power law distribution with a spectral index of 2-3. To explain the UHECR flux, these hadronic models usually assume the equipartition of the energy between electrons (or photons) and protons. Neutrinos and anti-neutrinos are produced in a ratio 1:2:0 (respectively for electron, muon and tau neutrinos) according to the delta resonance:

$$p + \gamma \rightarrow \Delta^+ \rightarrow \pi^+ + n$$
$$\pi^+ \rightarrow \mu^+ + \nu_\mu$$
$$\mu^+ \rightarrow e^+ + \nu_e + \text{anti-}\nu_\mu$$

Assuming that the secondary pions receive ~20% of the proton energy per interaction and each secondary lepton takes 25% of the pion energy, each flavour of neutrino is emitted with ~5% of the proton energy, dominantly in the TeV-PeV range.

This paper presents the derivation of the neutrino flux for individual GRBs from the observed electromagnetic data of the Swift satellite [5]. Then, using the simulated KM3NeT detector performance [3], the number of events is computed for each GRB below the horizon at the detector location with a measured redshift.

## 2. Prompt gamma-ray and neutrino spectra

The prompt GRB photon spectrum $N_\gamma$ is usually fitted using a broken power law introduced by Band et al [6]:

$$N_\gamma(E) \approx F_\gamma \times \begin{cases} (E)^\alpha & if\,(E < \varepsilon^b) \\ (\varepsilon^b)^{\alpha-\beta}(E)^\beta & if\,(E \geq \varepsilon^b) \end{cases}$$

The two spectral indices are usually scattered around average values of $\alpha_\gamma \sim -1$ and $\beta_\gamma \sim -2$. The value of the energy break is usually contained between 100 and 800 keV. The neutrino spectrum can be derived assuming that the proton spectrum follows the electron spectrum at the source [7]. The neutrino spectrum can then, in the first approximation, be described using the following formula:

$$E_\nu^2 \frac{dN}{dE_\nu} \approx \frac{F_\gamma f_\pi}{8\varepsilon_e \ln(10) T_{90}} \times \begin{cases} \left(\frac{E_\nu}{\varepsilon_\nu^b}\right)^{-\beta-1} & if\,(E_\nu < \varepsilon_\nu^b) \\ \left(\frac{E_\nu}{\varepsilon_\nu^b}\right)^{-\alpha-1} & if\,(\varepsilon_\nu^b < E_\nu < \varepsilon_\pi^b) \\ \left(\frac{E_\nu}{\varepsilon_\nu^b}\right)^{-\alpha-1}\left(\frac{E_\nu}{\varepsilon_\pi^b}\right)^{-2} & if\,(E_\nu > \varepsilon_\pi^b) \end{cases}$$

The normalization of the spectrum is given by the product of the gamma-ray fluence ($F_\gamma$) with the fraction of proton energy transferred to the pions, $f_\pi$ and also with the factor 1/8 (since half of the photohadronic interactions result in four neutrinos). The factor $1/\varepsilon_e$ accounts for the fraction of total energy in electrons compared to protons in the jet. T90 represents the time needed to accumulate 90% of the fluence. The detailed of the calculation can be found in [8].

## 3. Application to the Swift data

The BAT instrument [9] on board the Swift satellite has detected around 300 GRBs since December 2004. Of these, around half are below the horizon for the detector while one third have a measured redshift. These constraints leave a sample for this study containing 58 GRBs.

BAT can measure the gamma-ray spectrum only between 15 and 150 keV. This excludes the direct measurement of the energy break and the high energy spectral index for most of the GRBs. Based on the knowledge of BATSE [10], the spectrum is extrapolated up to few MeV. To do this, we fix the HE spectral index at $2.3^{+0.7}_{-0.3}$ and an energy break between 100 and 800 keV [21]. Using these parameters, a maximum and a minimum spectrum can be derived (Fig.1). From this, the gamma-ray fluence is globally corrected.

Figures 1 and 2 present respectively the prompt photon spectrum and the predicted neutrino spectrum for GRB071227 (z = 0.383, $F_\gamma$ = 2.2 $10^{-7}$ erg/cm², duration = 1.8s). The background spectrum (mainly from atmospheric neutrinos) is computed using the Volkova parameterisation [11] and a search window of 1.5°. Due to the large parameter error, the predicted spectrum has around one order of magnitude error. Above a 10 TeV energy threshold, the signal to background ratio has a maximum for the analysis of the GRB events.

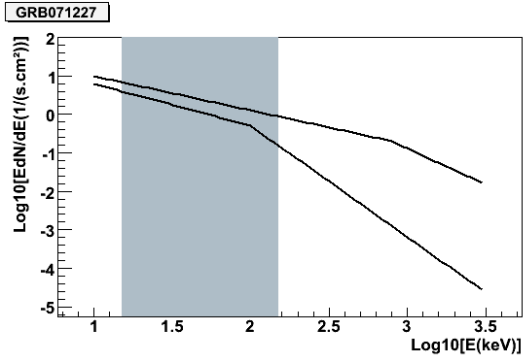

Fig. 1: Minimum and maximum prompt gamma-ray spectra for the GRB071227. The shaded region represents the detection capability of the BAT instrument [9].

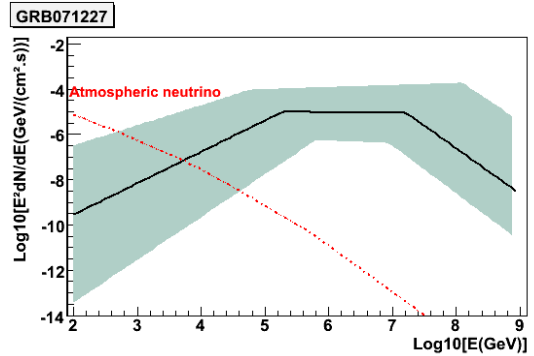

Fig. 2: Prediction of the neutrino spectrum for the GRB071227. The shaded region shows the incertitude. The descending dashed curve corresponds to the neutrino background flux assuming a search window of 1.5° (Volkova [11]).

## 4. Prediction for a KM3NeT detector

The number of events produced for each GRB is derived by convoluting the effective area of the KM3NeT detector and the predicted flux using a 10 TeV energy threshold. The effective area (Fig. 3) is obtained using a full Monte Carlo and the 'NESSY' reconstruction [12]. This effective area is also compared to the ANTARES and IceCube area. To take into account the atmospheric neutrino background, we assumed that a GRB is a point source and the angular resolution is 0.2°.

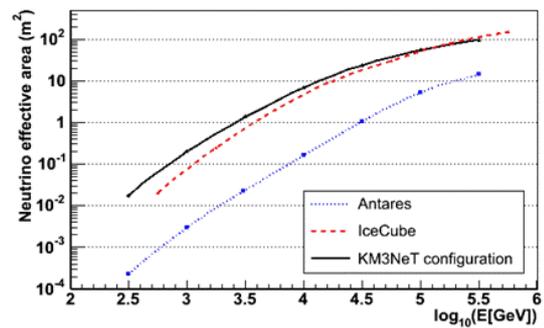

Fig. 3: Effective area for the KM3NeT detector (upper solid curve). The upper and lower dashed curves correspond respectively to the ANTARES [13] and IceCube [14] effective areas.

The two plots of figure 4 represent the distribution of signal and background for the 58 Swift GRBs. The number of GRB and background events is on average respectively ~$5\times10^{-3}$ and ~$6\times10^{-5}$ with plus or minus one order of magnitude. For very bright bursts, the number of events above 10 TeV can reach 0.05.

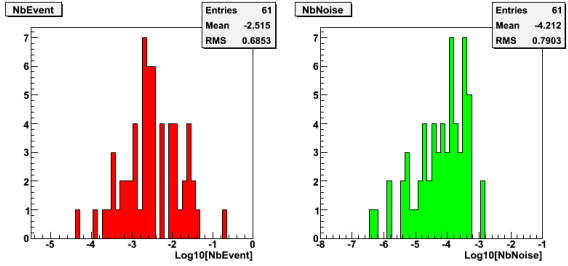

Fig. 4: Distribution of the number of detected neutrinos for GRB events (left plot) and background events (right plot) obtained after reconstruction with the KM3NeT detector for the 58 Swift GRBs. The background has been calculated using a search window of 1.5°.

Even though this number of events is quite low, we can easily determine a criterion to detect an individual burst of $10^{51}$ erg/s luminosity. For each distance or redshift, the gamma-ray fluence is recalculated. For different parameter values (α, T90…) the number of events is computed. In order to detect at least 2 events from the same burst in a km3-sized detector, the gamma-ray fluence of the burst has to be greater than ~ $4\ 10^{-4}$ erg/cm². In other word, all bursts closer than ~150 Mpc (or a redshift of ~0.04) can be detected. This result is compatible with previous studies [15]. The fluence criterion has been already satisfied by a few bursts detected by BATSE during its 10 years of operation. AMANDA has analysed its data and has made no positive GRB detection [16], [17].

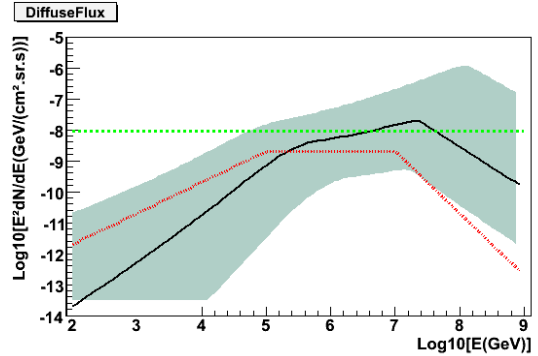

Fig. 5: Prediction for the contribution of GRB on the neutrino diffuse flux (ascending curve) with its uncertainty (shaded region). The dotted curve represents the prediction by Waxman and Bahcall [18]. Finally, the dashed flat line corresponds to the so-call Waxman Bahcall limit computed with the normalization of the neutrino flux to that of the UHECR [19].

By averaging all neutrino spectra for each Swift GRB and correcting for the BAT field of view (1.4 sr), we can easily compute the contribution of GRBs to the neutrino diffuse flux (Fig.5). This result is compatible with the prediction made by Waxman and Bahcall [18]. Due to the large spectrum-to-spectrum variability, it is important to take into account all the individual contributions to calculate a diffuse flux.

### 5. Summary

Among all the possible astrophysical sources, GRBs offer one of the most promising perspectives for the detection of cosmic neutrinos. Even the detection of a small number of neutrinos correlated with GRBs could prove without ambiguity the presence of hadronic acceleration to ultra high energy.

The main result of this study is that according to this model it is possible to individually detect GRBs with the forthcoming km$^3$ neutrino telescopes, IceCube or KM3NeT. This work is consistent with previous studies based on BATSE data [20]. Due to the large spectrum-to-spectrum variability, it is important to take into account all the individual contributions to calculate a diffuse flux.

Of course, this model has important limitations due to the assumptions made. For example, only protons are taken into account (not nuclei) and the proton spectrum, which has a direct impact on the neutrino flux, is usually renormalized to the electron spectrum. Moreover, many parameters, including the baryon load of the jet have large uncertainties. Expected fluxes vary considerably from one model to another; future observations derived directly from the data will help to constrain these models.

**Acknowledgements**

This work is supported through the EU-funded FP6 KM3NeT Design Study Contract N$^{o.}$ 011937.